\documentclass{iau} 
\usepackage{amsmath,amssymb}
\usepackage{graphicx}
\usepackage[version=4]{mhchem}

\newcommand{\phh}{{\rm p\mathchar`- H_{2}}}
\newcommand{\ohh}{{\rm o\mathchar`- H_{2}}}

\newcommand{\reacteq}[4]{{\rm #1} + {\rm #2} \rightleftharpoons {\rm #3} + {\rm #4}} 
\makeatletter 
\newcommand{\ion}[2]{#1$\;${\small\rmfamily\@Roman{#2}}\relax}
\makeatother

\title[Isotopic fractionation in interstellar molecules] %% give here short title %%
{Isotopic fractionation in interstellar molecules}

\author[Kenji Furuya]   %% give here short author list %%
{Kenji Furuya$^1$}
\affiliation{$^1$Center for Computational Sciences, University of Tsukuba, 1-1-1 Tennoudai, 305-8577 Tsukuba, Japan
\\email: {\tt furuya@ccs.tsukuba.ac.jp}}

\pubyear{2017}
\volume{332}  %% insert here IAU Symposium No.
\jname{Astrochemistry VII Through the Cosmos from Galaxies to Planets.}
\editors{Maria Cunningham, Tom Millar \& Yuri Aikawa, eds.}
\begin{document}

\maketitle

\begin{abstract}
The level of isotopic fractionation in molecules provides insights into their formation environments and how they formed.
In this article, we review hydrogen and nitrogen isotopic fractionation in low-mass star-forming regions.
Interstellar molecules are significantly enriched in deuterium.
The importance of the nuclear spin states of light species on deuterium fractionation and the usefulness of 
singly and doubly deuterated molecules as chemical tracers are discussed.
Observations have revealed that molecules in prestellar cores are enriched in or depleted in $^{15}$N depending on molecules.
Compared with deuterium fractionation chemistry, our understanding of $^{15}$N fractionation chemistry is not well established.
We briefly discuss potential $^{15}$N fractionation routes, 
i.e., isotopic-exchange reactions and isotopic selective photodissociation of N$_2$.
In addition, the selective freeze-out of $^{15}$N atoms onto dust grains around the transition between N atoms and \ce{N2} 
is discussed as a potential mechanism that causes the depletion of $^{15}$N in the gas phase.
%% add here a maximum of 10 keywords, to be taken form the file <Keywords.txt>
\end{abstract}

\firstsection % if your document starts with a section,
              % remove some space above using this command.
\section{Introduction}
More than 150 different molecules have been detected in the interstellar medium (ISM). 
One of the significant features in interstellar molecules is that their isotope ratios for a given element, 
such as the D/H ratio, the $^{14}$N/$^{15}$N ratio, and the $^{12}$C/$^{13}$C ratio, significantly 
(up to several orders of magnitude for hydrogen and up to a factor of a few for nitrogen and carbon)
deviate from their elemental abundance ratios.
This is called isotopic fractionation.
The level of isotopic fractionation in molecules depends on their formation environments and how they formed.
Then isotope ratios in molecules have been widely used as tracers of physical and chemical conditions in the ISM.
Primitive solar system materials, such as comets, also show isotopic fractionation 
(e.g., \cite[Mumma \& Charnley 2011]{mumma11}; \cite[Altwegg et al. 2015]{altwegg15}). 
By comparing isotope ratios in the interstellar materials and those in the solar system materials, 
their possible chemical link has been discussed 
(e.g., \cite[Caselli \& Ceccarelli 2012]{caselli12}; \cite[Cleeves et al. 2014]{cleeves14}; \cite[Furuya et al. 2017]{furuya17}).

Isotopic fractionation in the ISM can be driven by a couple of mechanisms: 
isotopic-exchange reactions in the gas phase, isotope selective UV photodissociation, and grain surface reactions.
In addition to the local processes, the elemental abundance ratios depend on the galactocentric distance and are not constant with time, 
because of stellar nucleosynthesis over time (e.g., \cite[Wilson 1999]{wilson99}).
Then one should be careful about differences in elemental abundances when comparing the isotope fractionation ratios in different objects, 
for example cometary material versus interstellar material.

In this paper, we briefly review deuterium fractionation and nitrogen isotope fractionation in low-mass star-forming regions.
Due to limitations of space, we mostly focus on basic processes of deuterium fractionation, including the nuclear spin chemistry of \ce{H2} (Section 2) 
and what information can be extracted from observationally derived deuterium fractionation ratios (Section 3).
Nitrogen isotope fractionation is discussed in Section 4.
More comprehensive reviews can be found in e.g., \cite[Caselli \& Ceccarelli (2012)]{caselli12}, \cite[Bergin (2014)]{bergin14},
\cite[Ceccarelli et al. (2014)]{ceccarelli14}, and \cite[Willacy et al. (2015)]{willacy15}.

\section{Deuterium fractionation chemistry in star-forming regions}

\subsection{Deuterium fractionation mechanisms}
Star formation takes place in molecular clouds.
Molecular clouds are formed from diffuse atomic gas by gravitational collapse or by accretion caused by shock waves 
(e.g., \cite[Dobbs et al. 2014]{dobbs14}, \cite[Inutsuka et al. 2015]{inutsuka15}).
During the formation of molecular clouds, chemical transition from atomic hydrogen to \ce{H2} occurs 
through recombination of H atoms on grain surfaces (e.g., \cite[Vidali  2013]{vidali13}).
The elemental abundance of deuterium with respect to hydrogen is [D/H]$_{\rm elem}$ = $1.5\times10^{-5}$ 
in the local ISM (\cite[Linsky 2003]{linsky03}).
Chemical transition of atomic deuterium to \ce{HD} occurs through the gas-phase reaction between \ce{D+} and \ce{H2} 
and the grain surface reaction between H and D atoms (\cite[Watson 1973]{watson73}). 
Because HD is much less abundant than \ce{H2}, HD needs a higher column density of the ISM gas for self-shielding.
Then the transition from D atoms to HD resides deeper into the cloud, compared to the \mbox{\ion{H}{1}}/\ce{H2} transition 
(\cite[van Dishoeck \& Black 1986]{vandishoeck86}).

In molecular clouds and prestellar cores, hydrogen and deuterium are primary in \ce{H2} and HD, respectively.
Deuterium fractionation can be understood as the process that distributes deuterium in HD to other species.
Deuterium fractionation in the gas phase is mainly triggered by isotope exchange reactions (\cite[e.g., Watson et al. 1976]{watson76});
\begin{align}
\reacteq{H_3^+}{HD}{H_2D^+}{H_2}.\label{eq:dfrac_reaction1}
\end{align}
At a typical temperature of molecular clouds, $\sim$10 K, the backward reaction is inefficient due to the endothermicity (but see below) and 
thus \ce{H2D+} becomes abundant with time with respect to \ce{H3+}.
Since \ce{H3+} plays a central role in the production of various gaseous molecules (\cite[Herbst \& Klemperer 1973]{herbst73}), 
the deuterium enrichment in \ce{H3+} is transfered to other gaseous molecules.
For example, the \ce{N2D+}/\ce{N2H+} abundance ratio is $>$0.01 in prestellar cores (e.g., \cite[Crapsi et al. 2005]{crapsi05}).
Even doubly and triply deuterated molecules, such as \ce{D2CO} and \ce{ND3}, have been detected 
(e.g., \cite[Bacmann et al. 2003]{bacmann03}, \cite[Roueff et al. 2005]{roueff05}).
With increasing temperature, the rate of the backward reaction becomes significant, and thus 
deuterium fractionation becomes less efficient.

\begin{figure}[t]
% \vspace*{-2.0 cm}
\begin{center}
 \includegraphics[width=4in]{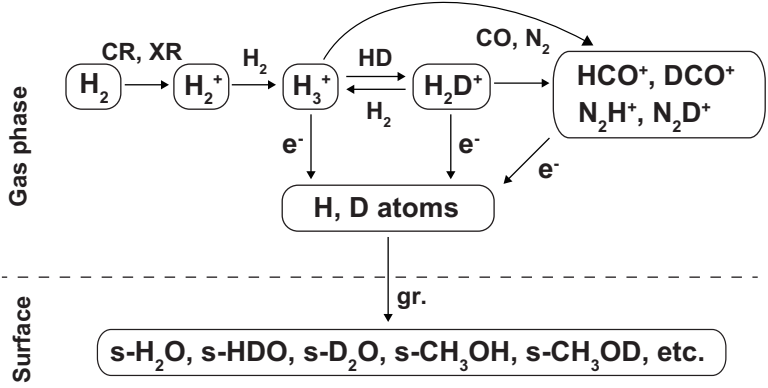} 
% \vspace*{-1.0 cm}
 \caption{Summary of the main deuterium fractionation pathways of gas and icy molecules driven by Reaction (\ref{eq:dfrac_reaction1}).
``gr.'' indicates adsorption of gas-phase species onto a grain surface followed by subsequent surface reactions.
``s-X'' indicates species X on a grain surface.
In extreme condition, where most neutrals are frozen out and the \ce{H2} OPR is lower than $\sim 6\times10^{-4}$,
multiply deuterated \ce{H3+}, \ce{D2H+} and \ce{D3+}, become more abundant than \ce{H3+}.
Multiply deuterated \ce{H3+} is not shown in this figure for simplicity (see e.g., \cite[Walmsley et al. 2004]{walmsley04} for more information).}
 \label{fig:dchem_network}
\end{center}
\end{figure}

The enrichment of deuterium in \ce{H3+} leads to the increased atomic D/H ratio in the gas phase, 
since D atoms are formed via dissociative recombination of deuterated \ce{H3+}.
The main components of interstellar ices, such as water, ammonia, methanol, and their deuterated forms, 
are formed via  sequential hydrogenation/deuteration of atoms and molecules on dust grain surfaces 
(\cite[Hama \& Watanabe 2013]{hama13}; \cite[Linnartz et al. 2015]{linnartz15} for recent reviews).
Then deuteration levels of icy molecules reflect the increased atomic D/H ratio in the gas phase 
(e.g., \cite[Tielens 1983]{tielens83}; \cite[Brown \& Millar 1989]{brown89}).
Summary of the deuterium fractionation chemistry is shown in Figure \ref{fig:dchem_network}.

On the other hand, \ce{H2D+} can further react with HD to form \ce{D2H+} and \ce{D3+} (e.g., \cite[Roberts et al. 2003]{roberts03}; \cite[Vastel et al. 2004]{vastel04});
\begin{align}
&\reacteq{H_2D^+}{HD}{D_2H^+}{H_2},\\
&\reacteq{D_2H^+}{HD}{D_3^+}{H_2},\label{eq:dfrac_reaction3}
\end{align}
Since \ce{D3+} is the end product of a series of Reactions (\ref{eq:dfrac_reaction1})-(\ref{eq:dfrac_reaction3}) and does not react with HD, 
\ce{D3+} can be more abundant than \ce{H3+} in extreme conditions (complete depletion of heavy elements, 
see e.g., \cite[Walmsley et al. 2004]{walmsley04} and \cite[Ceccarelli \& Dominik 2005]{ceccarelli05}).
Other deuterium isotopic exchange reactions, such as \ce{CH3+} + HD, become more important than the \ce{H2D+} chemistry 
at warm temperatures ($\gtrsim$50 K, e.g., \cite[Millar et al. 1989]{millar89}; \cite[Roueff et al. 2013]{roueff13}), 
because of their higher exothermicity.

\subsection{Ortho-to-para ratio of \ce{H2}}
\begin{figure}[t]
% \vspace*{-2.0 cm}
\begin{center}
 \includegraphics[width=2.5in]{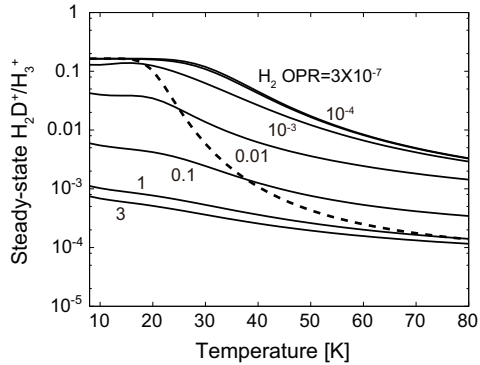} 
% \vspace*{-1.0 cm}
 \caption{The steady-state \ce{H2D+}/\ce{H3+} ratio at given \ce{H2} OPR in the range of $3\times 10^{-7}$-3 as functions of gas temperature 
(see Appendix of \cite[Furuya et al. 2015]{furuya15} for more information).
In the black dotted line, \ce{H2} OPR is assumed to be in thermal equilibrium.
CO and electron abundances with respect to hydrogen nuclei are set to be 10$^{-4}$ and 10$^{-8}$, respectively.
}
 \label{fig:h2dp}
\end{center}
\end{figure}

In addition to temperature, the level of deuterium fractionation depends on various parameters.
Freeze-out of neutral species, such as CO and atomic oxygen, onto dust grains (i.e., ice formation) enhances deuterium fractionation, 
because the neutral species are the main destroyers of deuterated \ce{H3+} (e.g., \cite[Lepp et al. 1987]{lepp87}).
The rate of Reaction (\ref{eq:dfrac_reaction1}) in the backward direction strongly depends on the ortho-to-para ratios (OPRs) of \ce{H2} and \ce{H2D+} 
(and also \ce{H3+}). 
\ce{H2} and \ce{H2D+} have two nuclear spin state isomers, the ortho-form and the para-form.
For the ground states, o-\ce{H2} and o-\ce{H2D+} have higher internal energy than their corresponding para forms by 170 K and 87 K, respectively,
which are much higher than the typical temperature in molecular clouds (10-20 K).
Since the high internal energy helps to overcome the endothermicity of the backward reaction,
deuterium fractionation becomes less efficient as the ortho forms become abundant
(e.g., \cite[Pagani et al. 1992]{pagani92}; \cite[Gerlich et al. 2002]{gerlich02}; \cite[Flower et al. 2006]{flower06}; \cite[Hugo et al. 2009]{hugo09}).
The left panel of Figure \ref{fig:h2dp} shows the steady-state \ce{H2D+}/\ce{H3+} ratio as functions of gas temperature, varying the \ce{H2} OPR.
The steady-state \ce{H2D+}/\ce{H3+} ratio is roughly in inverse proportion to the \ce{H2} OPR at given temperature.
It should also be noted that at $\lesssim$20 K,
\ce{H2} rather than CO is the key regulator of the \ce{H2D+} chemistry as long as OPR(\ce{H2}) $\gtrsim$ 40$x(\ce{CO})$, 
where $x(\ce{CO})$ is the CO abundance with respect to hydrogen nuclei (\cite[Furuya et al. 2015]{furuya15}).
Then detailed understanding of the \ce{H2} OPR is required for understanding the deuterium chemistry.

The \ce{H2} OPR in molecular clouds and cores is not well-constrained; it requires detailed physical and chemical modeling to estimate
the \ce{H2} OPR from observations of molecules other than \ce{H2} in cold clouds and cores.
Observational studies in literature consistently suggest the \ce{H2} OPR is lower than unity.
(e.g., \cite[Pagani et al. 2009]{pagani09}, \cite[2011]{pagani11}; \cite[Br{\"u}nken et al. 2014]{brunken14}; \cite[Le Gal et al. 2014]{legal14}).

Theoretically, the \ce{H2} OPR in the ISM is determined by the competition between the \ce{H2} formation on grain surfaces 
and the ortho-para spin conversion in the gas phase via the proton exchange reactions between \ce{H2} and \ce{H+} (or \ce{H3+})
(\cite[Gerlich 1990]{gerlich90}; \cite[Le Bourlot 1991]{lebourlot91}; \cite[Honvault et al. 2011]{honvault11}).
\ce{H2} molecules form on grain surfaces with the statistical OPR of three (\cite[Watanabe et al. 2010]{watanabe10}). 
The o-p conversion in the gas phase reduces the \ce{H2} OPR to be the thermal equilibrium (TE) value of $9\exp(-170/T_{\rm gas})$ (\cite[Gerlich 1990]{gerlich90}).
Laboratory experiments have found that the o-p spin conversion also occurs on grain surfaces in a laboratory timescale 
(\cite[Watanabe et al. 2010]{watanabe10}; \cite[Ueta et al. 2016]{ueta16}).
It is not trivial to quantify the effect of the o-p conversion on surfaces in the ISM, because of the uncertainties of the o-p conversion rates 
and the residence time of \ce{H2} on surfaces (see e.g., \cite[Hincelin et al. 2015]{hincelin15} for the latter).
We do not consider the o-p conversion on a surface in the following discussions.

In astrochemical simulations, it has been generally assumed that hydrogen is already locked in \ce{H2} at the beginning of the simulations, 
and the focus is placed on the subsequent molecular evolution.
In this approach, the initial \ce{H2} OPR is treated as a free parameter, due to the lack of observational constraints.
The resultant deuteration ratio of molecules in the simulations strongly depends on the initial \ce{H2} OPR, since the
time it takes for the \ce{H2} OPR to reach steady state is longer than the free-fall timescale and the freeze-out timescale 
(e.g., \cite[Flower et al. 2006]{flower06}; \cite[Kong et al. 2015]{kong15}).
\cite[Furuya et al. (2015)]{furuya15} studied the evolution of the \ce{H2} OPR in forming molecular clouds from diffuse atomic gas.
They found that at the \mbox{\ion{H}{1}}/\ce{H2} transition, the \ce{H2} OPR is already lower than the statistical value of three 
by a factor of 10 or even more, 
because the timescale of the o-p conversion in the gas phase is shorter than the timescale of \ce{H2} formation. 
The exact value, however, depends on physical environments, gas-phase abundances of heavy metals (e.g., sulfur), and the PAH chemistry.
More studies of the transition from atomic gas to \ce{H2} gas (i.e., molecular cloud formation) are required for better understanding of the initial \ce{H2} OPR.

Whatever the initial \ce{H2} OPR is, the \ce{H2} OPR approaches the steady-state value with time.
The steady-state \ce{H2} OPR is given as follows (\cite[Le Bourlot 1991]{lebourlot91}; \cite[Furuya et al. 2015]{furuya15}):
\begin{align}
{\rm OPR}(\ce{H2})_{\rm st} &= \frac{9\exp(-170/T_{\rm gas}) 
+ b_{\rm o}\tau_{\rm o\rightarrow p}/\tau_{\rm H_2}}{1 + (1-b_{\rm o})\tau_{\rm o\rightarrow p}/\tau_{\rm H_2}}, \label{eq:oprsteady}
%\frac{\num{\ohh}}{\num{\phh} } &= \frac{9\exp(-170.5/T_{\rm gas}) + b_{\rm o}\beta}{1 + (1-b_{\rm o})\beta}, \label{eq:oprsteady}\\
%\beta &= \tau_{\rm o\rightarrow p}/\tau_{\rm H_2}, \label{eq:beta2}
\end{align}
where $b_o$ is the branching ratio to form $\ohh$ for H$_2$ formation on grain surfaces (0.75, \cite[Watanabe et al. 2010]{watanabe10}).
$\tau_{\rm o \rightarrow p}$ is the spin conversion timescale from $\ohh$ to $\phh$ via the proton exchange reactions in the gas phase, 
while $\tau_{\rm H_2}$ is the formation timescale of H$_2$ on grain surfaces.
The important point here is that there is a critical temperature under which the steady state \ce{H2} OPR is 
larger than the TE value (see e.g., Fig. 1 of \cite[Faure et al. 2013]{faure13});
the critical temperature is 20 K for typical molecular clouds, where $\xi_{\rm CR}/n_{\rm gas}$ is around 10$^{-21}$ cm$^{3}$ s$^{-1}$, and 
it becomes lower with decreasing $\xi_{\rm CR}/n_{\rm gas}$.
Below the critical temperature, the steady state \ce{H2} OPR is given by $b_{o}\tau_{\rm o\rightarrow p}/\tau_{\rm H_2}$,
that is a function of $\xi_{\rm CR}/n_{\rm gas}$.
At $T_{\rm gas} = 10$ K and $\xi_{\rm CR}/n_{\rm gas} = 10^{-21}$ cm$^{3}$ s$^{-1}$,
the steady state \ce{H2} OPR is $\sim$10$^{-3}$, i.e., much higher than the TE value of $\sim$10$^{-7}$ (e.g., \cite[Flower et al. 2006]{flower06}).
Note that the enhanced \ce{H2} OPR affects the \ce{H2D+} OPR thorough the non-reactive collision between \ce{H2} and \ce{H2D+} 
(e.g., \cite[Pagani et al. 1992]{pagani92}; \cite[Gerlich et al. 2002]{gerlich02}).
This is the reason why the \ce{H2D+} OPR is a good tracer of the \ce{H2} OPR (\cite[Br{\"u}nken et al. 2014]{brunken14}; \cite[Harju et al. 2017]{harju17}).

\section{What can we learn from deuteration of icy molecules?}
In this section, we discuss what information can be extracted from deuteration of icy molecules.
The reason why we focus on ices is that bulk ice mantle compositions preserve the past physical and chemical conditions which the materials experienced.
Deuteration measurements of the ISM ices rely heavily on the gas observations toward inner warm ($\gtrsim$150 K) regions in the deeply embedded
protostars, where all ices have sublimated.
High temperature gas-phase chemistry after ice sublimation could alter the original composition, 
but the chemical timescale is much longer ($\sim$10$^5$ yr for the ionization rate of \ce{H2} of 10$^{-17}$ s$^{-1}$) 
than the infalling timescale of the warm gas onto the central star if the spherical collapse is assumed 
(e.g., \cite[Charnley et al. 1997]{charnley97}; \cite[Aikawa et al. 2012]{aikawa12}; \cite[Taquet et al. 2014]{taquet14}; \cite[Wakelam et al. 2014]{wakelam14}).
Then the composition of the warm gas is most likely dominated by the ice composition, which was established in the cold prestellar stages.

\subsection{Singly deuterated molecules}
Ice formation in star-forming regions can be roughly divided into two stages 
(i.e.,  ice mantles have two layered structure in terms of their molecular compositions) as revealed by infrared ice observations 
(e.g., \cite[Pontoppidan 2006]{pontoppidan06}; \cite[\"Oberg et al. 2011]{oberg11}); 
in the early stage, \ce{H2O}-dominated ice layers are formed.
In the later stage, at higher extinction and density, the catastrophic CO freeze out happens, and 
ice layers which consist of CO and its hydrogenated species (\ce{H2CO} and \ce{CH3OH}) are formed.

Observations toward deeply embedded low mass protostars have revealed that the level of methanol deuteration, in particular the \ce{CH3OD}/\ce{CH3OH} ratio, 
is much higher than the HDO/\ce{H2O} ratio in the warm ($\gtrsim$150 K) gas around the protostars, where ices have sublimated 
($\sim$10$^{-2}$ versus $\sim$10$^{-3}$; \cite[Parise et al. 2006]{parise06}; \cite[Coutens et al. 2014]{coutens14}; \cite[Persson et al. 2013]{persson13}, \cite[2014]{persson14}).
Since the deuteration levels of icy molecules reflect the atomic D/H ratio in the gas phase,
this trend indicates that deuterium fractionation processes are more efficient in the later stage of the ice formation, 
most likely due to CO freeze out and the drop of the \ce{H2} OPR 
(\cite[Cazaux et al. 2011]{cazaux11}; \cite[Taquet et al. 2012]{taquet12}).
Then, generally speaking, one can constrain the relative timing of the formation of icy molecules by comparing their D/H ratios.

Care is required when one applies the above discussion to other icy molecules, such as formaldehyde;
the D/H ratio of icy molecules can be altered after their formation.
Once water and ammonia are formed, they do not efficiently react with D atoms to be deuterated (\cite[Nagaoka et al. 2005]{nagaoka05}).
On the other hand, the deuteration levels of formaldehyde and methanol can be enhanced by the substitution and abstraction reactions of H and D atoms at low temperatures ($\sim$10 K)
after their formation (\cite[Watanabe \& Kouchi 2008]{watanabe08}; \cite[Hidaka et al. 2009]{hidaka09}).
As another example, methylamine (\ce{CH3NH2}) is also subject to the H/D substitution reactions (\cite[Oba et al. 2014]{oba14}).
Moreover the efficiency of the substitution and abstraction reactions depend on a functional group in molecules, 
e.g., effective in the CH$_3$- group of methanol and not in its OH- group.
Indeed, the observationally derived \ce{CH2DOH}/\ce{CH3OD} ratio is $\sim$10 in the warm gas of embedded low-mass protostars, 
contrasting with the expectation that the ratio is unity, when methanol ices are simply formed via addition reactions of H and D atoms on CO ice 
(\cite[Parise et al. 2006]{parise06}; \cite[Taquet et al. 2012]{taquet12}).
The \ce{CH2DOH}/\ce{CH3OD} ratio is close to unity in high-mass hot cores (\cite[Ratajczak et al. 2011]{ratajczak11}).
These observations likely indicate that the substitution and abstraction reactions are efficient in low-mass star-forming regions, but not the high-mass sources 
(\cite[Taquet et al. 2012]{taquet12}).

\begin{figure}[t]
% \vspace*{-2.0 cm}
\begin{center}
 \includegraphics[width=5in]{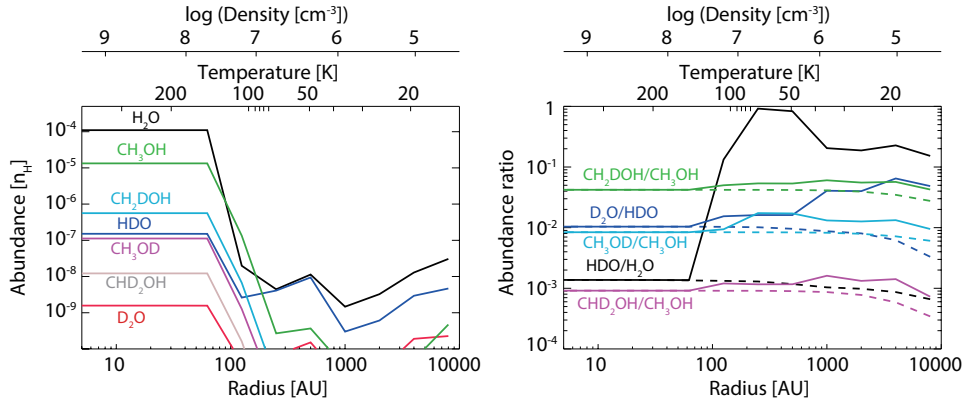} 
% \vspace*{-1.0 cm}
 \caption{Radial profile of the molecular abundances (left panel) and the abundance ratios of the molecules (right panel) 
as functions of distance form the central low-mass protostar in the deeply embedded stage 
(\cite[Furuya et al. 2016]{furuya16} with some updates).
The labels at the top represents the temperature and density structures.
The solid lines represent gaseous molecules, while the dashed lines represent icy molecules in the bulk ice
mantle.
}
 \label{fig:envelope}
\end{center}
\end{figure}

State-of-the-art astrochemical models, which take into account the dynamical evolution of gravitationally collapsing cores, the deuterium and nuclear spin state 
chemistry, and a layered ice structure, have reproduced these observations at least qualitatively (\cite[Taquet et al. 2014]{taquet14}; \cite[Furuya et al. 2016]{furuya16}).
Figure \ref{fig:envelope} shows radial profiles of the molecular abundances (left panel) and the abundance ratios of the molecules (right panel) 
in the deeply embedded protostellar stage as functions of distance from a central low-mass protostar.
Inside $\sim$60 AU from the central star, dust temperature is high enough to sublimate all ices ($\gtrsim$150 K), and the gas composition is determined by the ice sublimation.
At $>$150 K, the HDO/\ce{H2O} ratio is $\sim$10$^{-3}$ and \ce{CH3OD}/\ce{CH3OH} ratio is $\sim$10$^{-2}$, reproducing the observations.
The \ce{CH2DOH}/\ce{CH3OD} ratio is higher than unity due to the efficient abstraction reactions in the model, 
but the model still underestimates the ratio compared to the observationally derived value of $\sim$10.

It is worth noting that the gaseous HDO/\ce{H2O} ratio is decreased from the cold outer envelopes to 
the inner warm ($\gtrsim$150 K) regions by more than one order of magnitude,
which has been confirmed both by observations and astrochemical models 
(Fig. \ref{fig:envelope}; \cite[Coutens et al. 2012]{coutens12}, \cite[Persson et al. 2013]{persson13}).
Then interferometer observations are more suitable for the deuteration measurements of warm water than single dish observations, 
which provide the integrated emission from much larger spacial scales than the warm gas (\cite[J{\o}rgensen \& van Dishoeck 2010]{jorgensen10}).
According to astrochemical models, unlike water, methanol does not show the decrease of the D/H ratio from the outer envelopes to the inner warm regions,
because methanol efficiently forms only on grain surfaces and in the later stage of the ice formation (Fig. \ref{fig:envelope}; see also \cite[Taquet et al. 2014]{taquet14}).

\subsection{Doubly deuterated molecules}
In addition to singly deuterated molecules, doubly deuterated molecules in the inner warm regions around protostars 
can provide more constrains on the physical and chemical evolution in the cold prestellar stages.
Recently, \cite[Coutens et al. (2014)]{coutens14} found that 
the gaseous \ce{D2O}/HDO ratio is much higher than the HDO/\ce{H2O} ratio ($\sim$10$^{-2}$ versus $\sim$10$^{-3}$) 
in the inner warm regions of deeply embedded protostar NGC 1333-IRAS 2A.
This finding contrasts with the expectation that the ratio is 1/4 (i.e., the statistical ratio), 
when water ices are formed via addition reactions of H and D atoms with atomic oxygen  (\cite[Rodgers \& Charnley 2002]{rodgers02}).

This higher \ce{D2O}/HDO than HDO/\ce{H2O} can be understood by taking into account chemical evolution (see \cite[Furuya et al. 2016]{furuya16} for details).
Like the HDO/\ce{H2O} ratio, the \ce{D2O}/HDO ratio depends on the atomic D/H ratio.
Then the high \ce{D2O}/HDO ratio indicates that the water ice formation still continues when deuteration becomes more efficient due to 
CO freeze out and the drop of the \ce{H2} OPR.
Indeed, the \ce{D2O}/HDO ratio ($\sim$10$^{-2}$) is similar to the \ce{CH3OD}/\ce{CH3OH} ratio.
But, the amount of water ice formed in this late stage should be limited to keep the HDO/\ce{H2O} ratio low.
This requires that free atomic oxygen in the gas phase should be mostly ($>$90 \%) consumed before the epoch of CO freeze out.
The low abundance of atomic oxygen is consistent with the low abundance of \ce{O2} gas in a protostellar envelope (\cite[Y{\i}ld{\i}z et al. 2013]{yildiz13}).
This scenario was confirmed by the gas-ice astrochemical model of \cite[Furuya et al. (2016)]{furuya16} (see also Fig. \ref{fig:envelope}).
Then, multiple deuteration of water is a strong tool to constrain the past history of water ice formation 
and the amount of free oxygen that is available for water ice formation.

A similar idea can be applied to atomic nitrogen and ammonia ice.
The primary reservoir of nitrogen in dense star-forming clouds is not well constrained.
One of the potentially important reservoirs is atomic nitrogen (\cite[Maret et al. 2006]{maret06}),
but it is not directly observable in the cold gas.
By conducting astrochemical modelings of star-forming cores,
it was shown that if atomic nitrogen is the primary reservoir of nitrogen even when the catastrophic CO freeze out happens, 
the [\ce{ND2H}/\ce{NH2D}]/[\ce{NH2D}/\ce{NH3}] ratio is close to the statistical value of 1/3, 
whereas if atomic nitrogen is largely converted into gaseous \ce{N2} or icy molecules, the ratio should be significantly larger than 1/3
(Furuya \& Persson submitted).
The similar method could be applied to other elements, such as sulfur, by observing e.g., \ce{H2S}, \ce{H2CS} and their deuterated forms in the warm gas.

\begin{figure}[t]
% \vspace*{-2.0 cm}
\begin{center}
 \includegraphics[width=3.5in]{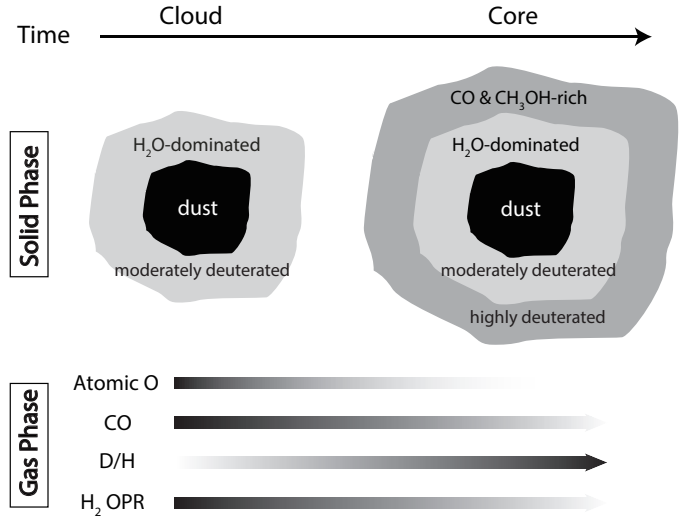} 
% \vspace*{-1.0 cm}
 \caption{Schematic view of gas and ice evolution in cold prestellar stages, clouds and cores.
Clouds) The main formation stage of \ce{H2O} ice. Deuteration in this stage is less efficient than in the later stage.
The majority of oxygen is locked in O-bearing molecules in this stage.
Cores) CO/\ce{CH3OH}-rich outer ice
layers are formed, while the formation of water ice continues with much
reduced efficiency, compared to the early stage.
Deuteration is more efficient, because of CO freeze out and the drop of the ortho-to-para ratio of \ce{H2}.
}
 \label{fig:ice_cartoon}
\end{center}
\end{figure}

Schematic view of gas and ice evolution in cold prestellar stages is summarized in Figure \ref{fig:ice_cartoon}.
The chemical and isotopic (at least for deuterium) compositions of the ISM ice are characterized not only by the bulk ice composition, 
but also by the differentiation within the ice mantle, reflecting the physical and chemical evolution during the prestellar stages.
There are several numerical methods to take into account a layered ice structure in astrochemical simulations, 
see e.g., \cite[Hasegawa \& Herbst (1993)]{hasegawa93}, \cite[Taquet et al. (2012)]{taquet12}, \cite[Vasyunin \& Herbst (2013)]{vasyunin13}, and \cite[Furuya et al. (2017)]{furuya17}.

\subsection{Caveat: H/D exchange reactions in warm ices}
Laboratory experiments have shown that thermally activated H-D exchanges between hydrogen-bonded molecules in mixed ices occur 
efficiently at warm temperatures of $\gtrsim$70 K before sublimation of polar ices 
(\cite[Ratajczak et al. 2009]{ratajczak09}; \cite[Faure et al. 2015b]{faure15b}; \cite[Lamberts et al. 2015]{lamberts15}).
The examples of this type of reaction are as follows:
\begin{align}
&\ce{H2O} + \ce{D2O} \rightleftharpoons \ce{2HDO}, \label{react:h2o+d2O} \\ 
&\ce{CH3OD} + \ce{H2O} \rightleftharpoons \ce{CH3OH} + \ce{HDO},
\end{align}
If these reactions are efficient, the original deuteration ratios established during the cold prestellar stages would be 
largely lost before ice sublimation (\cite[Faure et al. 2015a]{faure15a}).
It should be noted, however, that the H/D exchange occurs only between closely interacting molecules in ice,
and the efficiency of the H/D exchange in the inhomogenous ISM ice is unclear  (\cite[Ratajczak et al. 2009]{ratajczak09}; \cite[Faure et al. 2015a]{faure15a}).

\section{Nitrogen isotope fractionation in prestellar cores}
Nitrogen has two isotopes, $^{14}$N and $^{15}$N, with the elemental abundance ratio [$^{14}$N/$^{15}$N]$_{\rm elem}$ $\sim$ 300 
in the local ISM (\cite[Adande \& Ziurys 2012]{adande12}; \cite[Ritchey et al. 2015]{ritchey15}).
The dominant form of nitrogen in diffuse clouds is the atomic form with a small fraction of the molecular form
(\cite[Viala et al. 1986]{viala86}; \cite[Knauth et al. 2004]{knauth04}).
In dense molecular clouds, atomic N is (at least in part) converted into gaseous and icy molecules 
(e.g., \cite[Herbst \& Klemperer 1973]{herbst73}; \cite[Hily-Blant et al. 2010]{hilyblant10}),
but the primary form of nitrogen is not well constrained.
Gas-ice astrochemical models often predict most nitrogen is present as \ce{NH3} ice and \ce{N2} gas and ice 
(e.g., \cite[Daranlot et al. 2012]{daranlot12}; \cite[Chang \& Herbst 2014]{chang14}).

\begin{figure}[t]
% \vspace*{-2.0 cm}
\begin{center}
 \includegraphics[width=3in]{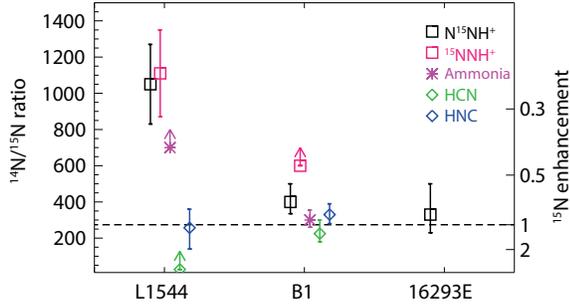} 
\caption{Observationally derived $^{15}$N fractionation in various molecules in prestellar cores, 
where the $^{15}$N isotopologues of \ce{N2H+} were observed 
(taken from \cite[Bizzocchi et al. 2013]{bizzocchi13}; \cite[Daniel et al. 2013]{daniel13}, \cite[2016]{daniel16}; 
\cite[Gerin et al. 2009]{gerin09}; \cite[Hily-Blant et al. 2013]{hilyblant13}; \cite[Milam \& Charnley 2012]{milam12}).
The N/$^{15}$N ratios in ammonia are the values for \ce{NH2D} in L1544 and for \ce{NH3} in Barnard 1.
The black dashed line indicates the elemental abundance ratio [N/$^{15}$N]$_{\rm elem}$ of $\sim$300 in the local ISM.
}
\label{fig:nchem_obs}
\end{center}
\end{figure}

Observations have revealed that N-bearing molecules in prestellar cores are enriched in or depleted in $^{15}$N depending on molecules.
Figure \ref{fig:nchem_obs} shows the observationally derived N/$^{15}$N ratio taken from literature. 
\ce{N2H+} and ammonia are either depleted in $^{15}$N or not fractionated depending on source, 
while nitriles are either enriched in $^{15}$N or not fractionated.
Note that some care is required for the interpretation of the enrichment of $^{15}$N in HCN and HNC as pointed out by \cite[Roueff et al. (2015)]{roueff15};
the N/$^{15}$N ratios of these molecules were derived from $^{13}$C species 
assuming the elemental $^{12}$C/$^{13}$C ratio.
HCN and HNC can be depleted in $^{13}$C by a factor of up to a few (see \cite[Langer et al. 1984]{langer84} for chemical fractionation of carbon isotope), 
and thus their N/$^{15}$N ratios could be underestimated by the same factor.

It has been argued that this observed trend reflects the different gas-phase formation pathways of these molecules;
gaseous ammonia and \ce{N2H+} are formed from \ce{N2}, while nitriles are formed from N atoms 
(\cite[Rodgers \& Charnley 2008]{rodgers08}; \cite[Wirstr\"om et al. 2012]{wirstrom12}; \cite[Hily-Blant et al. 2013]{hilyblant13}).
If so, then, the remaining question is what is the mechanism that causes the depletion of $^{15}$N in \ce{N2} and the enrichment of $^{15}$N in N atoms.

There are several studies of astrochemical modeling of $^{15}$N fractionation via isotope exchange reactions in prestellar cores 
(e.g., \cite[Terzieva \& Herbst 2000]{terzieva00}; \cite[Rodgers \& Charnley 2008]{rodgers08}; \cite[Wirstr\"om et al. 2012]{wirstrom12}).
The model of \cite[Wirstr\"om et al. (2012)]{wirstrom12} can explain the observed $^{15}$N fractionation in ammonia and nitriles, 
depending on the degree of CO freeze-out and the \ce{H2} OPR,
while their model predicts that \ce{N2H+} is enriched in $^{15}$N, being inconsistent with the observations.
Furthermore, recent quantum chemical calculations indicate the fractionation via isotope exchange reactions is not effective due to
the presence of activation barriers for some key reactions (\cite[Roueff et al. 2015]{roueff15}).
See contribution by E. S. Wirstr\"om in this volume for more details.

Another mechanism that can cause $^{15}$N fractionation is isotope selective photodissociation of \ce{N2} 
(\cite[Liang et al. 2007]{liang07}; \cite[Heays et al. 2014]{heays14}).
Photodissociation of \ce{N2} is subject to self-shielding.
Because N$^{15}$N is much less abundant than \ce{N2}, N$^{15}$N needs a higher column density of the ISM gas for self-shielding.
As a result, in some regions, N$^{15}$N is selectively photodissociated and then \ce{N2} is depleted in $^{15}$N, 
while atomic N is enriched in $^{15}$N.
\cite[Heays et al. (2014)]{heays14} developed a gas-phase astrochemical model of a molecular cloud, considering  
both isotope selective photodissociation of \ce{N2} and isotope exchange reactions for $^{15}$N fractionation.
They found that the isotope selective photodissociation is at work only in the limited regions, where the interstellar UV radiation field is not significantly attenuated 
and the transition from atomic nitrogen to \ce{N2} occurs.
In these regions, \ce{N2} and \ce{N2H+} are depleted in $^{15}$N, while nitriles are enriched in $^{15}$N.
Note, however, that prestellar cores typically have higher density ($>$10$^5$ cm$^{-3}$) and higher $A_V$
($>$10 mag for a dust continuum peak) compared to their cloud model.
Thus it seems difficult to explain the depletion of $^{15}$N in \ce{N2H+} in the framework of the present gas-phase chemical network.

On the other hand, the combination of gas and surface chemistry may cause the $^{15}$N fractionation.
A potentially important mechanism is the freeze out of N atoms followed by the conversion into less volatile species (e.g., \ce{NH3}) 
in the regions where N atoms are enriched in $^{15}$N, i.e., the \mbox{\ion{N}{1}}/\ce{N2} transition (Furuya \& Aikawa submitted; see also \cite[Bizzocchi et al. 2013]{bizzocchi13}).
In this case, $^{15}$N is selectively depleted in the gas phase with respect to $^{14}$N.
After the \mbox{\ion{N}{1}}/\ce{N2} transition, 
overall gas would be depleted in $^{15}$N compared to [N/$^{15}$N]$_{\rm elem}$, while ices would be enriched in $^{15}$N.
Once the differential N/$^{15}$N ratio between the gas-phase and the solid phase is established, it would remain 
as long as dust temperature is cold and ice sublimation is inefficient (until $\lesssim$150 K, if N atoms are mainly converted into ammonia ice).
The N/$^{15}$N ratios can vary among gaseous molecules, depending on the efficiency of  $^{15}$N fractionation by isotope exchange reactions.

\begin{figure}[t]
% \vspace*{-2.0 cm}
\begin{center}
 \includegraphics[width=4.8in]{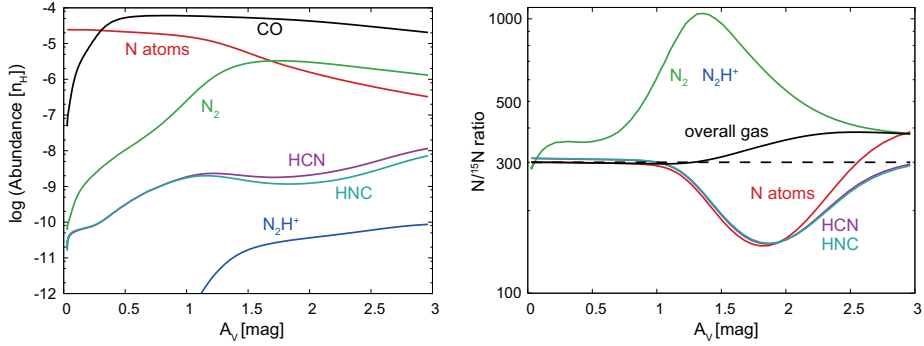} 
% \vspace*{-1.0 cm}
\caption{Molecular abundances of selected N-bearing molecules (left panel) and the N/$^{15}$N abundance ratios (right) 
as functions of visual extinction, $A_V$ (i.e., larger $A_V$ corresponds to the deeper regions into the cloud center) 
taken from a gas-ice astrochemical model of molecular cloud formation via converging flows.
The model includes gas phase chemistry, interaction between gas and grain, and grain surface chemistry in addition 
to isotope exchange reactions for $^{15}$N fractionation chemistry and isotope selective photodissociation of \ce{N2} (Furuya \& Aikawa submitted).
The black dashed line in the right panel indicates the elemental abundance ratio [N/$^{15}$N]$_{\rm elem}$ of 300,
while the black solid line indicates the N/$^{15}$N ratio in the gas phase.
}
 \label{fig:15n_model}
\end{center}
\end{figure}

In order to test this scenario, Furuya \& Aikawa (submitted) developed a gas-ice astrochemical model of a molecular cloud formation via converging flows of diffuse atomic gas, 
considering both the isotope selective photodissociation of \ce{N2} (\cite[Heays et al. 2014]{heays14}) and isotope exchange reactions for $^{15}$N fractionation (\cite[Roueff et al. 2015]{roueff15}).
Figure \ref{fig:15n_model} shows molecular abundances of N-bearing molecules in the model (left panel) and the N/$^{15}$N abundance ratios (right) 
as functions of visual extinction, $A_V$ (i.e., larger $A_V$ represents the deeper regions into the cloud center).
It was found that the isotope selective photodissociation is efficient only around the \mbox{\ion{N}{1}}/\ce{N2} transition, 
consistent with \cite[Heays et al. (2014)]{heays14} ($1\,\,{\rm mag} \lesssim A_V \lesssim 2\,\,{\rm mag}$ in this model). 
Around the \mbox{\ion{N}{1}}/\ce{N2} transition, \ce{N2H+} is significantly depleted in $^{15}$N, while nitriles are significantly enriched in $^{15}$N.
In deeper regions into the cloud, at $A_V \gtrsim 2$ mag, the overall gas is somewhat depleted in $^{15}$N due to the selective freeze-out of $^{15}$N followed by the $^{15}$\ce{NH3} formation 
on grain surfaces at $1\,\,{\rm mag} \lesssim A_V \lesssim 2\,\,{\rm mag}$.
As a consequence, \ce{N2H+} is somewhat depleted in $^{15}$N compared to [N/$^{15}$N]$_{\rm elem}$ even in the regions where interstellar UV radiation is significantly attenuated.
Nitriles have the similar N/$^{15}$N ratios to [N/$^{15}$N]$_{\rm elem}$ as a result of the local $^{15}$N fractionation chemistry via isotope exchange reactions.
The degree of $^{15}$N depletion in the overall gas by this mechanism depends on the amount of $^{15}$N atoms that are frozen out onto grain surfaces 
around the \mbox{\ion{N}{1}}/\ce{N2} transition.

%\begin{discussion}
%\end{discussion}


\begin{thebibliography}{}
\bibitem[Adande \& Ziurys (2012)]{adande12} Adande, G. R., \& Ziurys, L. M. 2012, \textit{ApJ}, 744, 194

\bibitem[Aikawa et al. (2012)]{aikawa12} Aikawa, Y., Wakelam, V., Hersant, F., Garrod, R., \& Herbst, E. 2012, \textit{ApJ}, 760, 40

\bibitem[Altwegg et al. (2015)]{altwegg15} Altwegg, K., Balsiger, H., Bar-Nun, A., et al. 2015, \textit{Science}, 347, 1261952

\bibitem[Bacmann et al. (2003)]{bacmann03} Bacmann A., Lefloch B., Ceccarelli C., Steinacker J., Castets A., Loinard L., 2003, \textit{ApJ}, 585, L55

\bibitem[Bergin (2014)]{bergin14} Bergin, E. A. 2014. Astrobiology: An astronomer's perspective. American Institute of Physics Conference Series 1638, 5-34

\bibitem[Bizzocchi et al. (2013)]{bizzocchi13} Bizzocchi, L., Caselli, P., Lenardo, E., \& Dore, L. 2013, \textit{A\&A}, 555, 109

\bibitem[Brown \& Millar (1989)]{brown89} Brown, P.D., \& Millar, T. J. 1989, \textit{MNRAS}, 237, 661

\bibitem[Br{\"u}nken et al. (2014)]{brunken14} Br{\"u}nken, S., Sipil{\"a}, O., Chambers, E. T. et al. 2014, \textit{Nature}, 516, 219

\bibitem[Caselli \& Ceccarelli (2012)]{caselli12} Caselli, P., \& Ceccarelli, C. 2012, \textit{ARA\&A}, 20, 56

\bibitem[Cazaux et al. (2011)]{cazaux11} Cazaux, S., Caselli, P., \& Spaans, M. 2011, \textit{ApJ}, 741, L34

\bibitem[Ceccarelli et al. (2014)]{ceccarelli14} Ceccarelli, C., Caselli, P., Bockel\'{e}e-Morvan, D., Mousis, O., Pizzarello, S., Robert, F. \& Semenov, D. 2014,  Protostars \& Planets VI (Tucson: Univ. Arizona Press), 859

\bibitem[Ceccarelli \& Dominik (2005)]{ceccarelli05} Ceccarelli, C., \& Dominik, C. 2005, \textit{A\&A}, 440, 583

\bibitem[Chang \& Herbst (2014)]{chang14} Chang, Q., \& Herbst, E. 2014, \textit{ApJ}, 787, 135

\bibitem[Charnley et al. (1997)]{charnley97} Charnley, S. B, Tielens, A. G. G. M, Rodgers, S. D. 1997, \textit{ApJ}, 482, 203 

\bibitem[Cleeves et al. (2014)]{cleeves14} Cleeves, L. I., Bergin, E. A., Alexander, C. M. O., et al. 2014, \textit{Science}, 345, 1590

\bibitem[Coutens et al. (2014)]{coutens14} Coutens, A., J{\o}rgensen, J. K., Persson, M., et al. 2014, \textit{ApJL}, 792, L5

\bibitem[Coutens et al. (2012)]{coutens12} Coutens, A., Vastel, C., \& Caux, E., et al. 2012, \textit{A\&A}, 539, 132

\bibitem[Crapsi et al. (2005)]{crapsi05} Crapsi, A., Caselli, P., Walmsley, C. M., et al. 2005, \textit{ApJ}, 619, 379

\bibitem[Dalgarno (2006)]{dalgarno06} Dalgarno, A. 2006, \textit{PNAS}, 103, 12269

\bibitem[Daniel et al. (2016)]{daniel16} Daniel, F., Faure, A., Pagani, L., et al. 2016, \textit{A\&A}, 592, 45

\bibitem[Daniel et al. (2013)]{daniel13} Daniel, F., G\'erin, M., Roueff, E., et al. 2013, \textit{A\&A}, 560, 3

\bibitem[Daranlot et al. (2012)]{daranlot12} Daranlot, J., Hincelin, U., Bergeat, A., et al. 2012, \textit{PNAS}, 109, 10233

\bibitem[Dobbs et al. (2014)]{dobbs14} Dobbs, C. L., Krumholz, M. R., Ballesteros-Paredes, J., et al. 2014, Protostars \& Planets VI, 3

\bibitem[Faure et al. (2015a)]{faure15a} Faure, A., Faure, M., Theul\'e, P., Quirico, E., \& Schmitt, B. 2015, \textit{A\&A}, 584, A98

\bibitem[Faure et al. (2015b)]{faure15b} Faure, M., Quirico, E., Faure, A., et al. 2015, \textit{Icarus}, 261, 14

\bibitem[Faure et al. (2013)]{faure13} Faure, A., Hily-Blant, P., Le Gal, R., Rist, C., \& Pineau des For$\hat{\rm e}$ts, G. 2013, \textit{ApJ}, 770, L2

\bibitem[Fedoseev et al. (2015)]{fedoseev15} Fedoseev, G., Ioppolo, S., \& Linnartz, H. 2015, \textit{MNRAS}, 446, 449

\bibitem[Flower et al. (2006)]{flower06} Flower, D. R., Pineau Des For$\hat{\rm e}$ts, G., Walmsley, C. M. 2006, \textit{A\&A}, 449, 621

\bibitem[Furuya et al. (2015)]{furuya15} Furuya, K., Aikawa, Y., Hincelin, U., Hassel, G. E., Bergin, E. A., Vasyunin, A. I., \& Herbst, E. 2015, \textit{A\&A}, 584, 124

\bibitem[Furuya et al. (2016)]{furuya16} Furuya, K., van Dishoeck, E. F., Aikawa, Y. 2016, \textit{A\&A}, 586, 127

\bibitem[Furuya et al. (2017)]{furuya17} Furuya, K., Drozdovskaya, M. N., Visser, R., van Dishoeck, E. F., Walsh, C., Harsono, D., Hincelin, U, Taquet, V. 2017, \textit{A\&A}, 599, 40

\bibitem[Gerin et al. (2009)]{gerin09} Gerin, M., Marcelino, N., Biver, N., et al. 2009, \textit{A\&A}, 498, L9

\bibitem[Gerlich (1990)]{gerlich90} Gerlich, D. J. 1990, \textit{Chem. Phys.}, 92, 2377

\bibitem[Gerlich et al. (2002)]{gerlich02} Gerlich, D., Herbst, E., \& Roueff, E. 2002, \textit{Plan. Sp. Sci.}, 50, 1275

\bibitem[Hama \& Watanabe (2013)]{hama13} Hama T., \& Watanabe N., 2013, \textit{Chem. Rev.}, 113, 8783

\bibitem[Hama et al. (2012)]{hama12} Hama, T., Kuwahata, K., Watanabe, N., et al. 2012, \textit{ApJ}, 757, 185

\bibitem[Harju et al. (2017)]{harju17} Harju, J., Sipil{\"a}, O., Br{\"u}nken, S., et al. 2017, ApJ, 840, 63

\bibitem[Hasegawa \& Herbst (1993)]{hasegawa93} Hasegawa, T. I., \& Herbst, E. 1993, \textit{MNRAS}, 263, 589

\bibitem[Heays et al. (2014)]{heays14} Heays, A. N., Visser, R., Gredel, R., et al. 2014, \textit{A\&A}, 562, 61

\bibitem[Herbst \& Klemperer (1973)]{herbst73} Herbst, E., \& Klemperer, W. 1973, \textit{ApJ}, 185, 505

\bibitem[Hidaka et al. (2009)]{hidaka09} Hidaka, H., Watanabe, M., Kouchi, A., \& Watanabe, N. 2009, \textit{ApJ}, 702, 291

\bibitem[Hily-Blant et al. (2013)]{hilyblant13} Hily-Blant, P., Bonal, L., Faure, A., \& Quirico, E. 2013, \textit{Icarus}, 223, 582

\bibitem[Hily-Blant et al. (2010)]{hilyblant10} Hily-Blant, P.,Walmsley, M., Pineau des Forêts, G., \& Flower, D. 2010, \textit{A\&A}, 513, 41

\bibitem[Hincelin et al. (2015)]{hincelin15} Hincelin, U., Chang, Q., \& Herbst, E. 2015, \textit{A\&A}, 574, 24

\bibitem[Honvault et al. (2011)]{honvault11} Honvault, P., Jorfi, M., Gonz\'alez-Lezana, T., Faure, A., \& Pagani, L. 2011, \textit{Phys. Rev. Lett.}, 107, 023201

\bibitem[Hugo et al. (2009)]{hugo09} Hugo, E., Asvany, O., \& Schlemmer, S. 2009, \textit{J. Chem. Phys.}, 130, 164302

\bibitem[Inutsuka et al. (2015)]{inutsuka15} Inutsuka, S.-I., Inoue, T., Iwasaki, K., \& Hosokawa, T. 2015, \textit{A\&A}, 580, A49

\bibitem[J{\o}rgensen \& van Dishoeck (2010)]{jorgensen10} J{\o}rgensen, J. K., \& van Dishoeck, E. F. 2010, \textit{ApJ}, 725, L172

\bibitem[Kalvans et al. (2017)]{kalvans17} Kalvans, J., Shmeld, I., Kalnin, J. R., Hocuk, S. 2017 \textit{MNRAS}, 467, 1763

\bibitem[Kong et al. (2015)]{kong15} Kong, S., Caselli, P., Tan, J. C., Wakelam, V., \& Sipil\"{a}, O. 2015, \textit{ApJ}, 804, 98

\bibitem[Knauth et al. (2004)]{knauth04} Knauth, D. C., Andersson, B.-G., McCandliss, S. R., \& Warren Moos, H. 2004, \textit{Nature}, 429, 636

\bibitem[Lamberts et al. (2015)]{lamberts15} Lamberts, T., Ioppolo, S., Cuppen, H. M., Fedoseev, G., \& Linnartz, H. 2015, \textit{MNRAS}, 448, 3820

\bibitem[Langer et al. (1984)]{langer84} Langer, W. D., Graedel, T. E., Frerking, M. A., \& Armentrout, P. B. 1984, \textit{ApJ}, 277, 581

\bibitem[Le Bourlot (1991)]{lebourlot91} Le Bourlot, J. 1991, \textit{A\&A}, 242, 235

\bibitem[Le Gal et al.  (2014)]{legal14} Le Gal, R., Hily-Blant, P., Faure, A., et al. 2014, \textit{A\&A}, 562, 83

\bibitem[Lepp et al. (1987)]{lepp87} Lepp, S., Dalgarno, A. \& Sternberg, A. 1987, \textit{ApJ}, 321, 383

\bibitem[Liang et al. (2007)]{liang07} Liang, M.-C., Heays, A. N., Lewis, B. R., Gibson, S. T., \& Yung, Y. L. 2007, \textit{ApJ}, 664, L115

\bibitem[Linnartz et al. (2015)]{linnartz15} Linnartz H., Ioppolo S., Fedoseev G., 2015, \textit{Int. Rev. Phys. Chem.}, 34, 205

\bibitem[Linsky (2003)]{linsky03} Linsky, J. L. 2003, \textit{Space Sci. Rev.}, 106, 49

\bibitem[Maret et al. (2006)]{maret06} Maret, S., Bergin, E. A., \& Lada, C. J. 2006, \textit{Nature}, 442, 425

\bibitem[Milam \& Charnley (2012)]{milam12} Milam, S. N. \& Charnley, S. B. 2012, in Lunar and Planetary Inst. Technical Report, Vol. 43, 
Lunar and Planetary Institute Science Conference Abstracts, 2618

\bibitem[Millar et al. (1989)]{millar89} Millar, T. J., Bennet, A., \& Herbst, E. 1989, \textit{ApJ}, 340, 906

\bibitem[Mumma \& Charnley (2011)]{mumma11} Mumma, M. J., \& Charnley, S. B. 2011, \textit{ARA\&A}, 49, 471

\bibitem[Nagaoka et al. (2005)]{nagaoka05} Nagaoka, A., Watanabe, \& N., Kouchi, A. 2005, \textit{ApJ}, 624, L29

\bibitem[Oba et al. (2012)]{oba12} Oba, Y., Watanabe, N., Hama, T., et al. 2012, \textit{ApJ}, 749, 67

\bibitem[Oba et al. (2014)]{oba14} Oba Y., Chigai T., Osamura Y., Watanabe N., Kouchi A., 2014, \textit{Meteor. Planet. Sci.}, 49, 117

\bibitem[\"Oberg et al. (2011)]{oberg11} \"Oberg, K. I., Boogert, A. C. A., Pontoppidan, K. M., et al. 2011, \textit{ApJ}, 740, 109

\bibitem[Pagani et al. (1992)]{pagani92} Pagani, L., Wannier, P. G., Frerking, M. A., et al. 1992, \textit{A\&A}, 258, 472

\bibitem[Pagani et al. (2009)]{pagani09} Pagani, L., Vastel, C., Hugo, E., et al. 2009, \textit{A\&A}, 494, 623

\bibitem[Pagani et al. (2011)]{pagani11} Pagani, L., Roueff, E., \& Lesaffre, P. 2011, \textit{A\&A}, 739, L35

\bibitem[Parise et al. (2006)]{parise06} Parise, B., Ceccarelli, C., Tielens, A. G. G. M., et al. 2006, \textit{A\&A}, 453, 949

\bibitem[Persson et al. (2013)]{persson13} Persson, M. V., J{\o}rgensen, J. K., \& van Dishoeck, E. F. 2013, \textit{A\&A}, 549, L3

\bibitem[Persson et al. (2014)]{persson14} Persson, M. V., J{\o}rgensen, J. K., van Dishoeck, E. \& Harsono, D. 2014, \textit{ApJ}, 563, 74

\bibitem[Pontoppidan (2006)]{pontoppidan06} Pontoppidan, K. M. 2006, \textit{A\&A}, 453, L47

\bibitem[Ratajczak et al. (2009)]{ratajczak09} Ratajczak, A., Quirico, E., Faure, A., Schmitt, B., \& Ceccarelli, C. 2009, \textit{A\&A}, 496, L21

\bibitem[Ratajczak et al. (2011)]{ratajczak11} Ratajczak, A., Taquet, V., Kahane, C., et al. 2011, \textit{A\&A}, 528, L13

\bibitem[Ritchey et al. (2015)]{ritchey15} Ritchey, A. M., Federman, S. R., \& Lambert, D. L. 2015, \textit{ApJL}, 804, L3

\bibitem[Roberts et al. (2003)]{roberts03} Roberts, H., Herbst, E., \& Millar, T. J. 2003, \textit{ApJ}, 591, L41

\bibitem[Rodgers \& Charnley (2002)]{rodgers02} Rodgers, S. D., \& Charnley, S. B. 2002, \textit{Planet. Space Sci.}, 50, 1125

\bibitem[Rodgers \& Charnley (2008)]{rodgers08} Rodgers, S. D., \& Charnley, S. B. 2008, \textit{ApJ}, 689, 1448

\bibitem[Roueff et al. (2013)]{roueff13} Roueff, E., Gerin, M., Lis, D. C., et al. 2013, \textit{Journal of Physical Chemistry A}, 117, 9959

\bibitem[Roueff et al. (2015)]{roueff15} Roueff, E., Loison, J. C., \& Hickson, K. M. 2015, \textit{A\&A}, 576, 99

\bibitem[Roueff et al. (2005)]{roueff05} Roueff, E., Lis, D. C., van der Tak, F. F. S., Gerin, M., \& Goldsmith, P. F. 2005, \textit{A\&A}, 438, 585

\bibitem[Taquet et al. (2012)]{taquet12} Taquet, V., Ceccarelli, C. \& Kahane, C. 2012, \textit{ApJ}, 748, L3

\bibitem[Taquet et al. (2014)]{taquet14} Taquet, V., Charnley, S. B., \& Sipil\"{a}, O. 2014, \textit{ApJ}, 791, 1

\bibitem[Taquet et al. (2013)]{taquet13} Taquet, V., Peters, P. S., Kahane, C., et al. 2013, \textit{A\&A}, 550, 127

\bibitem[Terzieva \& Herbst (2000)]{terzieva00} Terzieva, R., \& Herbst, E. 2000, \textit{MNRAS}, 317, 563

\bibitem[Tielens (1983)]{tielens83} Tielens, A. G. G. M. 1983, \textit{A\&A}, 119, 177

\bibitem[Ueta et al. (2016)]{ueta16} Ueta, H., Watanabe, N., Hama, T., Kouchi, A., 2016, \textit{Phys. Rev. Lett.}, 116, 253201

\bibitem[van Dishoeck \& Black (1986)]{vandishoeck86} van Dishoeck, E. F., \& Black, J. H. 1986, \textit{ApJS}, 62, 109

\bibitem[Vastel et al. (2004)]{vastel04} Vastel, C., Phillips, T. G., \& Yoshida, H. 2004, \textit{ApJ}, 606, L127

\bibitem[Vasyunin \& Herbst (2013)]{vasyunin13} Vasyunin, A. I., \& Herbst, E. 2013, \textit{ApJ}, 762, 86

\bibitem[Viala et al. (1986)]{viala86} Viala, Y. P. 1986, \textit{A\&AS}, 64, 391

\bibitem[Vidali (2013)]{vidali13} Vidali, G. 2013, ChRv, 113, 8752

\bibitem[Wakelam et al. (2014)]{wakelam14} Wakelam, V., Vastel, C., Aikawa, Y., et al. 2014, \textit{MNRAS}, 445, 2854

\bibitem[Walmsley et al. (2004)]{walmsley04} Walmsley, C. M., Flower, D. R., \& Pineau Des For$\hat{\rm e}$ts, G. 2004, \textit{A\&A}, 418, 1035

\bibitem[Watanabe \& Kouchi (2008)]{watanabe08} Watanabe, N., \& Kouchi, A. 2008, \textit{Prog. Surf. Sci.}, 83, 439

\bibitem[Watanabe et al. (2010)]{watanabe10} Watanabe, N., Kimura, Y., Kouchi, A., Chigai, T., Hama, T., \& Pirronello, V. 2010, \textit{ApJ}, 714, L233

\bibitem[Watson (1973)]{watson73} Watson, W. D. 1973, \textit{ApJL}, 183, L17

\bibitem[Watson et al. (1976)]{watson76} Watson, W. D. 1976, \textit{Reviews of Modern Physics}, 48, 513

\bibitem[Willacy et al. (2015)]{willacy15} Willacy, K., Alexander, C., Ali-Dib, M., et al. 2015, SSRv, 197, 151

\bibitem[Wilson (1999)]{wilson99} Wilson, T. L. 1999, Reports on Progress in Physics, 62, 143

\bibitem[Wirstr\"om et al. (2012)]{wirstrom12} Wirstr\"om, E. S., Charnley, S. B., Cordiner, M. A., \& Milam, S. N. 2012, \textit{ApJ}, 757, L11

\bibitem[Y{\i}ld{\i}z et al. (2013)]{yildiz13} Y{\i}ld{\i}z, U. A., Acharyya, K., Goldsmith, P. F. et al. 2013, \textit{A\&A}, 558, 58

\end{thebibliography}
\end{document}